\documentclass{nature}

\usepackage{graphicx}
\usepackage{amsmath}
\usepackage{amssymb}
\usepackage{color}
\usepackage{dcolumn}
\usepackage{epsfig}
\usepackage{bm}
\usepackage[urlcolor=blue]{hyperref}
\hypersetup{backref, colorlinks=true, linkcolor=blue, citecolor=blue}

\title{Superconductivity of Cs$_3$C$_{60}$ at atmosphere pressure}

\author{Di Peng$^{1}$, Ren-Shu Wang$^{2,1}$, Li-Na Zong$^{1}$ \& Xiao-Jia Chen$^{1,2\ast}$}

\begin{document}

\maketitle
\begin{affiliations}
\item{Center for High Pressure Science and Technology Advanced Research, Shanghai, 201203, China}
\item{School of Science, Harbin Institute of Technology, Shenzhen 518055, China}\\
$^{\ast}$e-mail: xjchen2@gmail.com
\end{affiliations}

\vspace{0.2cm}
\begin{abstract}
Pressure as a clean and efficient tool can bring about unexpected extraordinary physical and chemical properties of matters\cite{mao}. The recent discoveries of superconductivity at nearly room temperature in hydrides\cite{droz,zulu} highlight the power of pressure in this aspect. Capturing such high critical temperature ($T_{c}$) superconductivity at atmosphere pressure for the technological applications is highly desired. The large-scale growth of diamond through the chemical vapor deposition\cite{liang} away from the usual high-pressure and high-temperature conditions\cite{diamond} fuels such a hope. Similar to hydrides, Cs-doped C$_{60}$\cite{07P,01G,02T,03G} was also found to exhibit superconductivity by the application of pressure with a comparable $T_c$ of 40 K as MgB$_2$\cite{naga}. Here, we report the successful realization of superconductivity in Cs-doped C$_{60}$ at atmosphere pressure. The phase is characterized to have the primitive cubic structure in the space group of $Pa\overline{3}$ with the stoichiometry of Cs$_3$C$_{60}$. The superconductivity is evidenced from the observations of both the Meissner effect and zero-resistance state. Although the pressure effects on superconductivity are different for the newly discovered Cs$_{3}$C$_{60}$ compared to the known two phases with the face-centered-cubic ($fcc$) and A15 structure\cite{01G,02T,03G}, the evolution of $T_c$ with the volume occupied per C$_{60}^{3-}$ anion for all these superconductors follows the same universal trend, suggesting the same pairing mechanism of the superconductivity. Such a trend together with the nearly linear $T_c$ vs the lattice constant in the structure with smaller unit-cell volumes and the neighbouring antiferromagnetic state in the structure with larger unit-cell volumes invites the electron-phonon coupling and the electron correlations together to account for the superconductivity in Cs$_3$C$_{60}$. The present results and findings suggest a new route to capturing the superconductivity which takes place at high pressures to atmosphere pressure environment.  
\end{abstract}

The discovery of superconductivity with $T_c$ of 18-19 K in K-doped C$_{60}$\cite{06H} and 28-29 K in Rb-doped C$_{60}$\cite{ross} is an important event in modern physical science based on the new form of carbon called `buckminsterfullerene' or in short fullerene\cite{krot}. These amazing molecules exhibit rather rich physical properties\cite{rmp1,rmp2} in their $fcc$ superconducting phase in the formula of A$_3$C$_{60}$ with triple valence for alkali metal (A)\cite{ste,mcc,had,kuz}. Although Cs shares a similar outer electronic structure with K and Rb, synthesizing the Cs$_3$C$_{60}$ superconducting phase turns out to be extremely difficult. The trace superconductivity with a maximum $T_c$ of 40 K at pressure of around 1.2 GPa was reported in Cs-doped C$_{60}$ sample\cite{07P}. However, the structure and composition of the superconducting phase can not be determined due to the poor sample quality and extremely small shielding fractions (SFs). Till 2008, bulk superconductivity with a maximum $T_c$ of 38 K at pressure of around 0.7 GPa was reported in a Cs-doped C$_{60}$ sample\cite{01G}  with SF of 67\%. Structural analysis\cite{01G}  revealed the sample includes 77.7(6)\% of a cubic phase with the A15 structure in the space group $Pm\overline{3}n$, 13.4(1)\% of a body-centered-orthorhomic ($bco$) phase, and 8.9(2)\% of a second $fcc$ phase in the space group $Fm\overline{3}m$. The comparison measurements supported the observed $T_c$ from the A15 Cs$_3$C$_{60}$ phase\cite{01G}. The superconductivity of this A15 phase was found to exhibit a dome-shape with pressure or volume occupied per fulleride anion\cite{01G}. Later, the electronic ground state of A15 Cs$_3$C$_{60}$ was determined to be an antiferromagnetic insulator with the Neel temperature ($T_N$) of $\sim$46 K at ambient pressure\cite{02T}. Applying pressure suppresses this antiferromagnetic state and induces superconductivity followed by a dome-shaped $T_c$ behaviour with a narrow regime for coexisting two states. All these features\cite{03G} were also observed in the samples containing 86\% $fcc$ Cs$_3$C$_{60}$ and A15 phase as low as 3\%. Compared to A15 Cs$_3$C$_{60}$\cite{01G,02T}, $fcc$ Cs$_3$C$_{60}$\cite{03G} was found to possess weak antiferromagnetism with a much lower $T_N$ value of 2.2 K and become superconducting easier at modest lower pressure but with a lower maximum $T_c$ of 35 K and larger SFs over the studied pressure range. Despite these differences, the dome-like $T_c$ behaviours for both A15 Cs$_3$C$_{60}$\cite{01G,02T} and $fcc$ Cs$_3$C$_{60}$\cite{03G} can be well scaled by the ratio of the on-site interelectron repulsion $U$ to the bare conduction bandwidth $W$, suggesting the importance of electron correlations and the same pairing mechanism of superconductivity. So far, the reported superconductivity was based solely on the magnetization measurements with the detected Meissner effect in A15 Cs$_3$C$_{60}$\cite{01G,02T} and $fcc$ Cs$_3$C$_{60}$\cite{03G}. However, the key character of the zero-resistance state for the superconductivity in Cs$_3$C$_{60}$ molecules is still missing and waits to be filled up over a decade. It is also interesting to explore whether the superconductivity can take place at atmosphere pressure in Cs-doped C$_{60}$ molecules. Finding answers to all these issues is the target of the present study.    

\noindent\textbf{Superconductivity from the Meissner effect and zero-resistivity state}

Cs-doped C$_{60}$ samples were synthesized by using the developed wet-chemical method\cite{wang,zong,wang1}, as described in Methods. Meissner effect and zero-resistance state are the essential evidence for superconductivity. We performed $dc$ and $ac$ magnetic susceptibility ($\chi$) measurements on the synthesized samples. Figure 1\textbf{a} shows the temperature-dependent $dc$ $\chi$ in the zero-field cooling (ZFC) and field-cooling (FC) runs with an applied magnetic field ($H$) of 5 Oe. The Meissner diamagnetic effect can be seen by the rapid decline of both curves at 21 K, where the temperature is defined as $T_c$. Both the ZFC and FC $\chi$ curves show a clear bifurcation, which is usually caused by flux pinning. The SF value can be enhanced significantly by optimizing the preparation conditions (Extended Data Fig. 1). Repeated annealing was found to be effective in increasing SF for this material (Extended Data Fig. 2). The right inset in Fig. 1\textbf{a} shows the $H$-dependent magnetization ($M$) at various temperatures of interest with the applied fields up to $\pm$7 T. The diamond-shaped $M$-$H$ curves illustrate a large hysteresis effect, which is typical for a type-II superconductor. As temperature is increased, the $M$-$H$ curve gradually shrinks inward near $T_c$ and eventually disappears at and above $T_c$. 

Figure 1\textbf{b} shows the the real part ($\chi^{\prime}_{ac}$) and imaginary component ($\chi^{\prime\prime}_{ac}$) of the $ac$ magnetic susceptibility. The $\chi^{\prime}_{ac}$ signal reflects the magnetic shielding effect, similar to the ZFC result of the $dc$ susceptibility. The $\chi^{\prime\prime}_{ac}$ part is the signal of the induced vortex current, which usually occurs only when the resistance is small and drop sharply such as the occurrence of the superconducting transition\cite{gom}. Therefore, the peak in the $\chi^{\prime\prime}_{ac}$ curve can be regarded as the signature of the zero-resistance state. The $\chi^{\prime}_{ac}$ and $\chi^{\prime\prime}_{ac}$ behaviours for the samples prepared with different synthesis conditions are summarized in Extended Data Fig. 3. With the increase of SF, the magnetic shielding signal in $\chi^{\prime}_{ac}$ becomes sharp, the $\chi^{\prime\prime}_{ac}$ peak gradually appears. It should be mentioned that the sharp peak in the $\chi^{\prime\prime}_{ac}$ vs. temperature curve is the first observation from the $ac$ $\chi$ for Cs-doped C$_{60}$, which has not been reported in the early studies\cite{01G,02T,03G}.  

Electrical transport measurements can provide direct evidence for the zero-resistance state. For $fcc$ or A15 Cs$_3$C$_{60}$, the experimental evidence for the superconductivity was mainly from the magnetic measurements and the electrical transport measurements have not been reported yet\cite{01G,02T,03G}. Figure 1\textbf{c} shows the temperature-dependent resistivity ($\rho$) of our synthesized Cs-doped C$_{60}$ with the highest SF under the applied magnetic fields up to 9 T. Note that the resistivity begins to drop sharply around 20 K and quickly reaches the zero-resistance state. As the magnetic field is increased, the zero-resistance transition shifts to low temperatures, indicating the suppression of superconductivity by the application of the magnetic field. By examining the resistivity behaviours for samples with various SF grown at different synthesis conditions (Extended Data Fig. 4), one learns that the higher the SF for the better quality of the sample, the narrower and sharper the resistance transition. Now the experimental evidence for the superconductivity from the zero-resistance state in the synthesized samples of Cs-doped C$_{60}$ is firmly established. This is also the first zero-resistance state ever observed for Cs-doped C$_{60}$\cite{01G,02T,03G}. We thus complete the essential Meissner effect and zero-resistance state for the discovered Cs-doped C$_{60}$ superconductors.

\noindent\textbf{Phase identification from the structural and vibrational properties}

The phase behaviour and structure character of Cs-doped C$_{60}$ were determined by combining the x-ray diffraction and Raman scattering measurements. The obtained diffraction profile together with the refined results is shown in Fig. 2\textbf{a}. Rietveld refinement analysis reveals that the synthesized material crystallizes in an orientationally-ordered primitive cubic form (space group $Pa\overline{3}$) with a stoichiometry of Cs$_3$C$_{60}$. The refined parameters are summarized in Extended Data Tables 1 and 2. The excellent refinements give the lattice constant $a$ of 14.259(18) \AA\/ for the studied Cs$_3$C$_{60}$. This value is in between $\sim$ 11.8 \AA\/ (for body-centered cubic $bcc$ Cs$_3$C$_{60}$)\cite{01G} and $\sim$ 14.8 \AA\/ (face-centered cubic $fcc$ Cs$_3$C$_{60}$)\cite{03G}. The large difference in the lattice constants also reveal the new phase of Cs$_3$C$_{60}$ at atmosphere pressure. The obtained $a$ value in the present study is comparable to that of $fcc$ K$_3$C$_{60}$ ($a\approx$14.2 \AA)\cite{wang1,ste} and Rb$_3$C$_{60}$ ($a\approx$14.5 \AA)\cite{mcc}. Thus, it is not surprising to observe superconductivity in these fullerides due to their close lattice constants.

Raman spectroscopy has been widely used in doped fullerides\cite{had,kuz} and other organic systems\cite{huang,wang0,yan,hg} for determining the number of transferred charges. The Raman spectra of the pristine and Cs-doped C$_{60}$ are shown in Fig. 2\textbf{b}. The parent C$_{60}$ possesses two $A_g$ and eight $H_g$ vibrational modes, consistent with the previous studies\cite{kuz}. Among these Raman active modes, only the two $A_g$ modes and the two lowest frequency $H_g$ modes have the relatively high scattering intensities after doping, as illustrated in Fig. 2\textbf{b}. The $A_g$(2) mode represents the pentagonal breathing and is often used to monitor the doped electrons\cite{wang,zong,wang1}. Upon Cs doping, the $A_g$(2) mode shifts from 1458 cm$^{-1}$ to 1440 cm$^{-1}$. The redshift of 18 cm$^{-1}$ corresponds to the transferred three electrons based on the empirical law of 6 cm$^{-1}$ per electron\cite{had}, consistent with the formula of Cs$_3$C$_{60}$ determined from our above structure study.

\noindent\textbf{Carrier character from the Hall effect}

The charge carrier character of newly synthesized Cs$_3$C$_{60}$ superconductor was evaluated in terms of the Hall effect measurements. This information is very necessary for the understanding of the superconducting properties themselves and also important for the theory development. Figure 3\textbf{a} shows $H$-dependent Hall resistivity ($\rho_{xy}$) in the normal state at the temperature range of 30-290 K. It can be seen that the $\rho_{xy}$-$H$ curve exhibits a linear behavior. This enables to obtain the Hall coefficient $R_H$ easily just through the linear fitting to the collected data points for each temperature. The temperature dependence of $R_H$ and the carrier concentration $n_H$ is thus obtained (Figs. 3\textbf{b} and 3\textbf{c}). Similar to the observation in $fcc$ K$_3$C$_{60}$\cite{wang}, $R_H$ changes sign around 250 K, indicating the existence of both the electron and hole conduction due to half-filling\cite{erw}. The temperature-dependent $n_H$ behaviour demonstrates that the major charge carriers of this superconductor are holes at room temperature but become electrons at low temperatures below 250 K. In addition, both $R_H$ and $n_H$ abruptly change around 110 K, implying a possible electronic structure change. Interestingly, the pristine C$_{60}$ was found to undergo a glass transition at around 90 K\cite{gug}. The similar transitions in the materials with the same structure suggest the possible same origin for these anomalies. The two anomalies at temperature of 110 and 250 K are also reflected from the resistivity behaviour (the $d\rho/dT$-$T$ curve), as shown in the insets of Extended Data Fig. 6. Here, the reported carrier character is the first for Cs-doped C$_{60}$ superconductors\cite{01G,02T,03G} and thus hopes to be used to understand the superconductivity in these molecules. 

\noindent\textbf{Determination of the superconducting parameters}

The upper critical field, lower critical field, coherence length, and penetration depth are the key parameters for the understanding of the superconducting properties and the superconductivity nature. The high quality of our samples ensures the resistivity and magnetization measurements at desired temperature and magnetic field conditions to determine such important parameters. 

The temperature-dependent resistivity measurements for a given magnetic field were used to determine the upper critical field $H_{c2}$ (see Methods). The measured results for Cs$_{3}$C$_{60}$ are summarized in Fig. 1\textbf{c}. For each applied magnetic field regarded as $H_{c2}$, we can have the corresponding $T_{c}$. The results for the pairs of $H_{c2}$ and $T_{c}$ are summarized in Fig. 1\textbf{d}. These $H_{c2}$ and $T_{c}$ data points can be well described by Ginzburg-Landau theory\cite{tin} (see Methods), yielding $H_{c2}$(0) of 63.4$\pm$1.7 T at the absolute zero temperature. This is the first $H_{c2}$(0) value ever reported for Cs-doped C$_{60}$ superconductors\cite{01G,02T,03G}. The obtained $H_{c2}$(0) value for Cs$_{3}$C$_{60}$ is significantly larger than 17-49 T reported for  K$_3$C$_{60}$\cite{holc,hou,kas}, and comparable to 57 T for Rb$_3$C$_{60}$ and 67 T for Rb$_2$CsC$_{60}$ but lower than 85 T for RbCs$_2$C$_{60}$\cite{kas}.
The high $H_{c2}$(0) values of fullerene superconductors might result from the orientational disorder between adjacent C$_{60}$ molecules. This advantage points out the importance of these superconductors for the potential technological applications.

Magnetic field-dependent magnetization measurements at different given temperatures were used to determine the lower critical field $H_{c1}$. The results for Cs$_{3}$C$_{60}$ in the superconducting state are shown in Extended Data Fig. 5\textbf{a}. Taking the magnetic field at the point where the $M$-$H$ curve deviates from the linearity as $H_{c1}$ for a given temperature, we obtained the temperature dependence of $H_{c1}$ (Extended Data Fig. 5\textbf{b}). The zero-temperature $H_{c1}$(0) of 30.4$\pm$0.7 Oe is thus obtained from the empirical fitting\cite{tin} to the obtained data points. The $H_{c1}$(0) value of this new superconductor is comparable to 42$\pm$1 Oe for K$_3$C$_{60}$ and 32$\pm$1 Oe for Rb$_3$C$_{60}$ synthesized from large single-crystals\cite{iro}.

Using the expressions $H_{c1}$(0) = ($\Phi_0$/4$\pi$$\lambda$$^2_L$)($\ln$$\lambda_L$/$\xi_{GL}$)\cite{tin} and $H_{c2}$(0) = $\Phi_0$/2$\pi$$\xi$$^2_{GL}$ with the flux quantum $\Phi_0$ = 2.0678$\times$10$^{-15}$ Wb, we obtain the zero-temperature Ginzburg-Landau coherence length $\xi$$_{GL}$ = 22.8$\pm$0.4 {\AA} and London penetration depth $\lambda_L$ = 5444$\pm$69 {\AA} with the help of the obtained $H_{c1}$(0) and $H_{c2}$(0) values. Thus, the Ginzburg-Landau parameter $\kappa$ = $\lambda_L$/$\xi_{GL}$ = 238.8$\pm$7.2 is obtained, further implying that Cs$_3$C$_{60}$ is a type-II superconductor. It is clearly seen that the obtained $\xi$$_{GL}$ is just 1.5 larger than $a$ in Cs$_3$C$_{60}$ and comparable to the short coherence lengths of cuprate superconductors.

\noindent\textbf{Pressure effect on superconductivity}

High-pressure resistivity measurements were carried out to draw the $T_{c}$ evolution for $Pa\overline{3}$ Cs$_3$C$_{60}$. Figure 4\textbf{a} shows the normalized $\rho$-$T$ curves in the temperature range of 1.8-25 K at various pressures up to 2.7 GPa. As pressure is increased, the superconducting transition shows a shift toward low temperatures. $T_c$ was found to decrease systematically with increasing pressure (Fig. 4\textbf{b}), yielding the initial pressure derivative of $T_c$, $dT_{c}$/$dP$=-10.0$\pm$0.4 K/GPa. The absolute value of $dT_{c}$/$dP$ for $Pa\overline{3}$ Cs$_3$C$_{60}$ is obviously larger than those reported for $fcc$ K$_3$C$_{60}$\cite{wang1,45S} and Rb$_3$C$_{60}$\cite{sparn1,die}, indicating its large pressure effect on superconductivity. The negative $dT_{c}$/$dP$ is totally different to the pressure-induced superconductivity followed by the huge enhancement reported in $fcc$ and A15 Cs$_3$C$_{60}$\cite{01G,02T,03G}. Thus, the different pressure effects on superconductivity classify $Pa\overline{3}$ Cs$_3$C$_{60}$ again as a new superconducting phase. In addition, the obtained nearly linear $T_c$-$P$ behavior is distinctly different from the dome-shaped phase diagram observed in $fcc$ and A15 Cs$_3$C$_{60}$\cite{01G,02T,03G} but similar to those reported for $fcc$ K$_3$C$_{60}$ and Rb$_3$C$_{60}$\cite{wang1,zho}. 

\noindent\textbf{Evolution of the structure with pressure}

Synchrotron x-ray powder diffraction data were collected to study the structural behaviour at high pressures. All the diffraction patterns in the studied pressure range remain the same compared to those at the ambient pressure (Fig. 4\textbf{c}). This indicates that the material keeps the same structure feature at high pressures as the initial one. No structural transformation can be detected from the collected data. Upon compression, these patterns shift to the high angles, indicating the lattice shrinkage. Fitting the patterns by using the same structure model with the space group $Pa\overline{3}$ as the initial phase turns out to be in good agreement. The obtained unit cell volume $V$ versus pressure is plotted in Fig. 4\textbf{d}. These experimental data points can be well described by the Murnaghan equation of state\cite{47M}, yielding the unit cell volume at ambient pressure $V_0$ of 2904.3$\pm$5.2 \AA\/$^3$, the bulk modulus $B_0$ of 19.0$\pm$1.5 GPa, and its pressure derivative $B^\prime_0$ of 8.6$\pm$1.1. The obtained $V_0$ is consistent with that determined from the structure data at ambient pressure (Fig. 2\textbf{a}), indicating the reliability of the high-pressure structural analysis. The obtained $B_0$ of 19.0$\pm$1.5 GPa for Cs$_3$C$_{60}$ is significantly larger than that for A15 Cs$_3$C$_{60}$ (16.4 GPa)\cite{01G} and $fcc$ Cs$_3$C$_{60}$ (13.7 GPa)\cite{03G}, but is comparable to those for K$_3$C$_{60}$ (17.3 GPa)\cite{wang1} and Rb$_3$C$_{60}$ (17.35 GPa and 20.5 GPa)\cite{die,lud}. Having these parameters, one can convert the $T_c$-$P$ data to the desired $T_c$-$a$ curve (Fig. 4\textbf{e}). It can be seen that $T_c$ decreases monotonically with the gradual reduction of $a$. This behavior is distinctly different from the dome-shaped $T_c$ phase diagram observed in other two Cs$_3$C$_{60}$ phases ($fcc$ and A15)\cite{01G,02T,03G}, but similar to those of K$_3$C$_{60}$\cite{wang1} and Rb$_3$C$_{60}$\cite{zho,die,lud}. The different $T_c$-$a$ relations compared to other Cs-doped C$_{60}$\cite{01G,02T,03G} further demonstrate that this synthesized Cs$_3$C$_{60}$ is a new superconducting phase. The similar $T_c$-$a$ relations in $Pa\overline{3}$ Cs$_3$C$_{60}$ compared to $fcc$ K$_3$C$_{60}$ and Rb$_3$C$_{60}$ are possibly a result of their similar C$_{60}^{3-}$ anion packings.

\noindent\textbf{Universal phase diagram of superconductivity}

The volume occupied per C$_{60}^{3-}$ anion ($V_{C_{60}^{3-}}$) is usually used as a measure to study fullerides across different systems, which might provide the clue for the microscopic mechanism of superconductivity. Putting the obtained $T_c$ and $V_{C_{60}^{3-}}$ data points in Fig. 5 for the newly discovered Cs$_3$C$_{60}$ superconductor with other superconducting phases\cite{01G,02T,03G}, one can find the beautiful evolution tendency for all the three superconducting phases of Cs$_3$C$_{60}$. The material is in the antiferromagnetic state at large volume, becomes superconducting upon slight compression with the suppression of the antiferromagnetic order, and possesses the maximum $T_{c}$ at an optimal volume (thus pressure) before its $T_{c}$ value declining upon further heavy lattice compression. The same trend held for three superconducting phases suggests the same pairing mechanism for them. It is interesting to notice that Cs$_3$C$_{60}$ exhibits the nearly linear $T_c$ and $a$ (or $V_{C_{60}^{3-}}$) relation in the \emph{Pa$\overline{3}$} structure (Fig. 4\textbf{e} and Fig. 5), the antiferromagnetic Mott insulating state, and the dome-shaped $T_c$-$V_{C_{60}^{3-}}$ behaviour (Fig. 5) in the $fcc$ and A15 structure\cite{01G,02T,03G}. The nearly linear $T_c$-$a$ relation is the indicator for the important electron-phonon interaction\cite{fle,zho}. The emergence of superconductivity by suppressing the neighbouring antiferromagnetic order highlights the dominant contribution of the electron correlations to the superconductivity\cite{01G,02T,03G,capone}. Thus, the beautiful and universal $T_c$-$V_{C_{60}^{3-}}$ behaviour in Cs$_3$C$_{60}$ obviously invites the electron-phonon interaction and electron correlations to work together responsible for the superconductivity\cite{rmp1,rmp2}. This also indicates that the theoretical considerations by putting both together\cite{rmp1,rmp2,han,nom} are favorable for fullerene superconductors.

\begin{addendum}

\item[Acknowledgements] The work at HPSTAR was supported by the National Key R\&D Program of China (Grant No. 2018YFA0305900). The work at HIT was supported by the Shenzhen Science and Technology Program (Grant No. KQTD20200820113045081) and the Basic Research Program of Shenzhen (Grant No. JCYJ20200109112810241). 

\end{addendum}

\newpage
\vspace{0.3cm}
\noindent\textbf{Extended Data Table 1 $\mid$ Refined parameters for the primitive cubic (space group \emph{Pa$\overline{3}$}) Cs$_3$C$_{60}$ phase obtained from the Rietveld analysis of synchrotron x-ray ($\lambda$ = 0.6199 \AA) powder diffraction data collected at room temperature.} The weighted profile and expected \emph{R}-factors are \emph{R$_{wp}$} = 3.161 $\%$ and \emph{R$_{exp}$} = 6.431 $\%$. Goodness-of-fit and \emph{R}-pattern are \emph{Gof} = 0.491 and \emph{R$_{p}$} = 2.533 $\%$. The lattice constant and unit cell volume are \emph{a} = 14.259(18) \AA\/ and \emph{V} = 2899.14(11) \AA$^3$. Errors for the last two digits are given in parentheses. The fractional occupancy ($N$) was fixed to 1 for both cesium and carbon atoms. The atomic position is represented by ($x$,$y$,$z$).
\begin{center}
	\vspace{0.01cm}
	\begin{tabular}{c c c c c }
		\hline
		\hline
        Atom &  $x$/$a$   &   $y$/$b$   &   $z$/$c$   &   $N$  \\
		\hline
		Cs(1)  &  0.5  &  0.5  &  0.5  &  1.0  \\
		Cs(2)  & 0.25290  &   0.25290   &  0.25290  &  1.0  \\
		C(1)   & 0.22900  &  -0.03200   &  0.10100  &  1.0  \\
		C(2)   & 0.24400  &  -0.05400   &  0.00600  &  1.0  \\
		C(3)   & 0.20600  &   0.06500   &  0.12900  &  1.0  \\
		C(4)   & 0.20700  &  -0.14000   & -0.03600  &  1.0  \\
		C(5)   & 0.17100  &  -0.09600   &  0.15900  &  1.0  \\
		C(6)   & 0.22400  &   0.11200   & -0.03700  &  1.0  \\
		C(7)   & 0.24400  &   0.01900   & -0.06400  &  1.0  \\
		C(8)   & 0.20500  &   0.13500   &  0.06200  &  1.0  \\
		C(9)   & 0.15000  &  -0.20200   &  0.02000  &  1.0  \\
		C(10)  & 0.13200  &  -0.17900   &  0.11900  &  1.0  \\
		\hline
		\hline
	\end{tabular}
\end{center}

\newpage
\vspace{0.3cm}
\noindent\textbf{Extended Data Table 2 $\mid$ Refined anisotropic thermal parameters (U$_{ij}$) for the primitive cubic (space group \emph{Pa$\overline{3}$}) Cs$_3$C$_{60}$.} Anisotropic thermal motion is described by six parameters of the diagonal and off diagonal terms of a three-by-three matrix.
\begin{center}
	\vspace{0.1cm}
	\begin{tabular}{c c c c c c c }
		\hline
		\hline
        Atom &  $U_{11}$   &   $U_{22}$   &   $U_{33}$   &   $U_{12}$   &  $U_{13}$   &  $U_{23}$  \\
		\hline
		Cs(1)  &  0.514(33)  &  0.514(33)  &  0.514(33)  &  -0.047(23)  &  -0.047(23)  &  -0.047(23)  \\
		Cs(2)  &  2.13(18)  &  2.13(18)  &  2.13(18)  &  -0.40(19)  &  -0.40(19)  &  -0.40(19)  \\
		C(1)   &  0.011(96)  &  0.50(15)  &  0.38(18)  &  0.56(12)  &  0.046(89)  &  0.16(12)  \\
		C(2)   &  0.00(19)  &  0.88(35)  &  2.55(50)  &  -0.75(23)  &  -0.40(20)  &  -0.38(28)  \\
		C(3)   &  0.000(83)  &  0.61(14)  &  1.36(20)  &  -0.72(11)  &  -0.84(11)  &  1.02(15)  \\
		C(4)   &  0.00(17)  &  1.32(52)  &  0.48(24)  &  0.76(26)  &  0.22(11)  &  0.13(22)  \\
		C(5)   &  5.2(6)  &  0.52(29)  &  6.9(2)  &  -1.26(87)  &  4.3(6)  &  -2.04(92)  \\
		C(6)   &  0.17(18)  &  0.13(16)  &  0.81(30)  &  -0.10(16)  &  0.74(21)  &  -0.12(24)  \\
		C(7)   &  2.1(1)  &  2.0(1)  &  1.54(94)  &  2.4(1)  &  -0.96(81)  &  -0.34(41)  \\
		C(8)   &  2.56(55)  &  0.03(11)  &  0.01(14)  &  0.76(19)  &  -0.69(25)  &  -0.307(98)  \\
		C(9)   &  2.66(81)  &  0.50(19)  &  0.50(21)  &  1.21(40)  &  0.71(37)  &  0.41(18)  \\
		C(10)  &  1.49(34)  &  2.40(58)  &  0.00(14)  &  -1.76(37)  &  0.91(23)  &  -1.03(30)  \\
		\hline
		\hline
	\end{tabular}
\end{center}

\newpage
\begin{center}
\includegraphics[width=0.95\columnwidth]{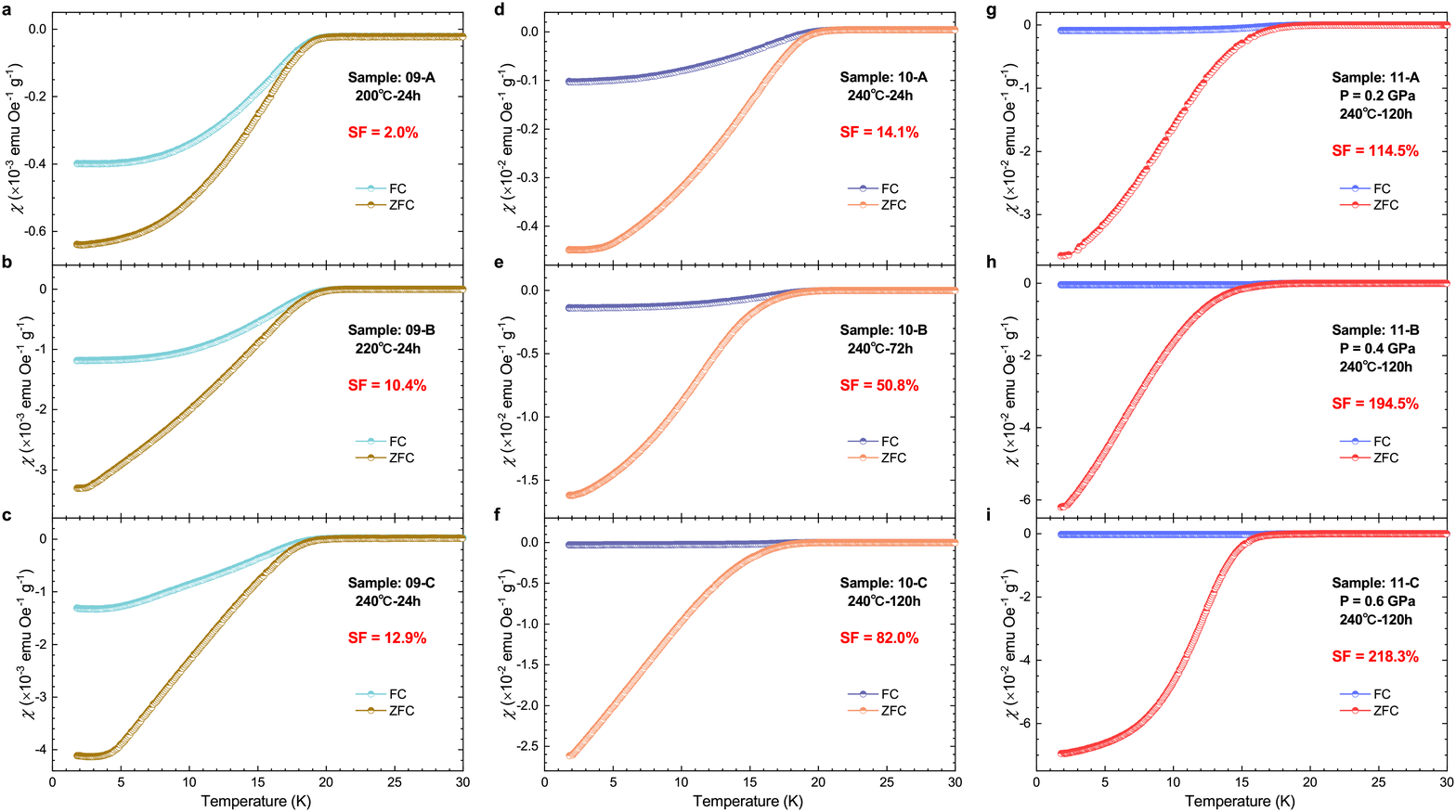}
\end{center}
\vspace{-1.0cm}
\noindent\textbf{Extended Data Figure 1 $\mid$ Evolution of the magnetic susceptibility ($\chi$) of Cs$_3$C$_{60}$ with samples at different synthesis conditions.} \textbf{a}, \textbf{b} and \textbf{c}, Temperature dependence of $\chi$ measured in both the zero-field-cooling (ZFC) and field-cooling (FC) runs of sample No. 9. Parts A, B and C of sample No. 9 were annealed at different temperatures of 200 $^\circ$C, 220 $^\circ$C and 240 $^\circ$C for 24 hours, respectively. The obtained superconducting shielding fraction (SF) increases from 2.0 $\%$ to 12.9 $\%$ with increasing temperature. There is a huge increase in SF from 200 $^\circ$C to 220 $^\circ$C, but the relatively small increase between 220 $^\circ$C and 240 $^\circ$C suggests that annealing the sample at 240 $^\circ$C is a relatively suitable temperature. \textbf{d}, \textbf{e} and \textbf{f}, Parts A, B and C of sample No. 10 were annealed at 240 $^\circ$C for periods ranging from 1 to 5 days, respectively. With increasing annealing time, SF increases substantially to 82.0 $\%$, indicating that increasing the time is effective in obtaining better samples. \textbf{g}, \textbf{h} and \textbf{i}, Parts A, B and C of sample No. 11 were annealed at 240 $^\circ$C for 5 days after the powder samples were pre-pressed into solid cylinders using different pressures. The SF values of the pre-pressed samples exceed 100 $\%$ and increase substantially with increasing pressure. Note that SF can exceed 100 $\%$ in the magnetic susceptibility measurements for the bulk samples and the effect of demagnetizing factor is not considered here.

\newpage
\begin{center}
\includegraphics[width=0.95\columnwidth]{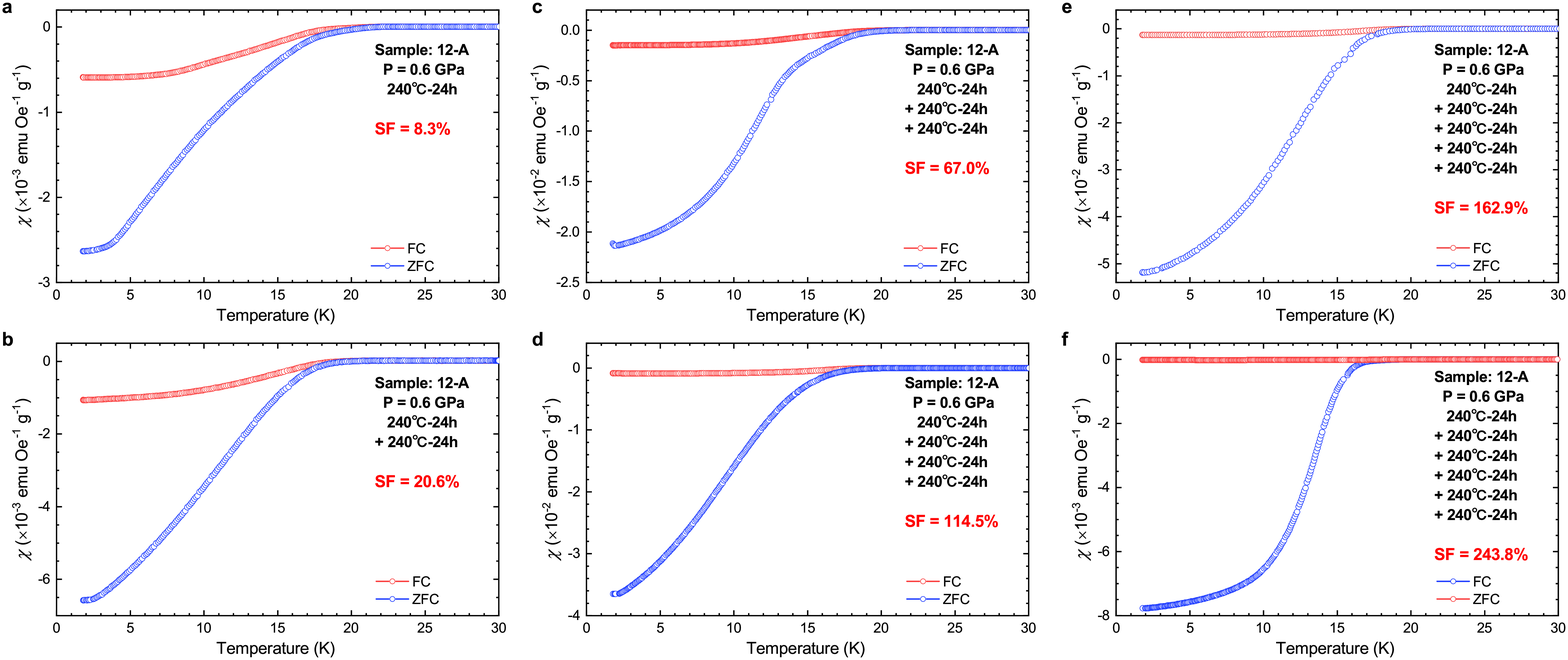}
\end{center}
\vspace{-1.3cm}
\noindent\textbf{Extended Data Figure 2 $\mid$ Evolution of the magnetic susceptibility ($\chi$) of Cs$_3$C$_{60}$ with the change of the annealing time to have the different superconducting shielding fraction (SF) for the same sample.} \textbf{a}, Temperature dependence of $\chi$ measured in both the ZFC and FC runs with the applied magnetic field of 10 Oe in the sweep mode. The superconducting phase appears after the sample was annealed at 240 $^\circ$C for 24 hours. The obtained SF is 8.3 $\%$. \textbf{b}, SF is increased to 20.6 $\%$ after an additional annealing of 24 hours. \textbf{c}, After adding another round of the annealing at 240 degrees Celsius for 24 hours, SF increases to 67.0 $\%$. \textbf{d}, After four times of the annealing at 240 degrees Celsius for a total of 84 hours, SF increases to 114.5 $\%$. Note that SF can exceed 100 $\%$ in the magnetic susceptibility measurements for the bulk samples. \textbf{e}, After the fourth time of the annealing at 240 $^\circ$C for 24 hours, SF continues to increase to 162.9 $\%$. \textbf{e}, Finally, SF increases to 243.8 $\%$ after annealing for a total of 132 hours.

\newpage
\begin{center}
\includegraphics[width=0.95\columnwidth]{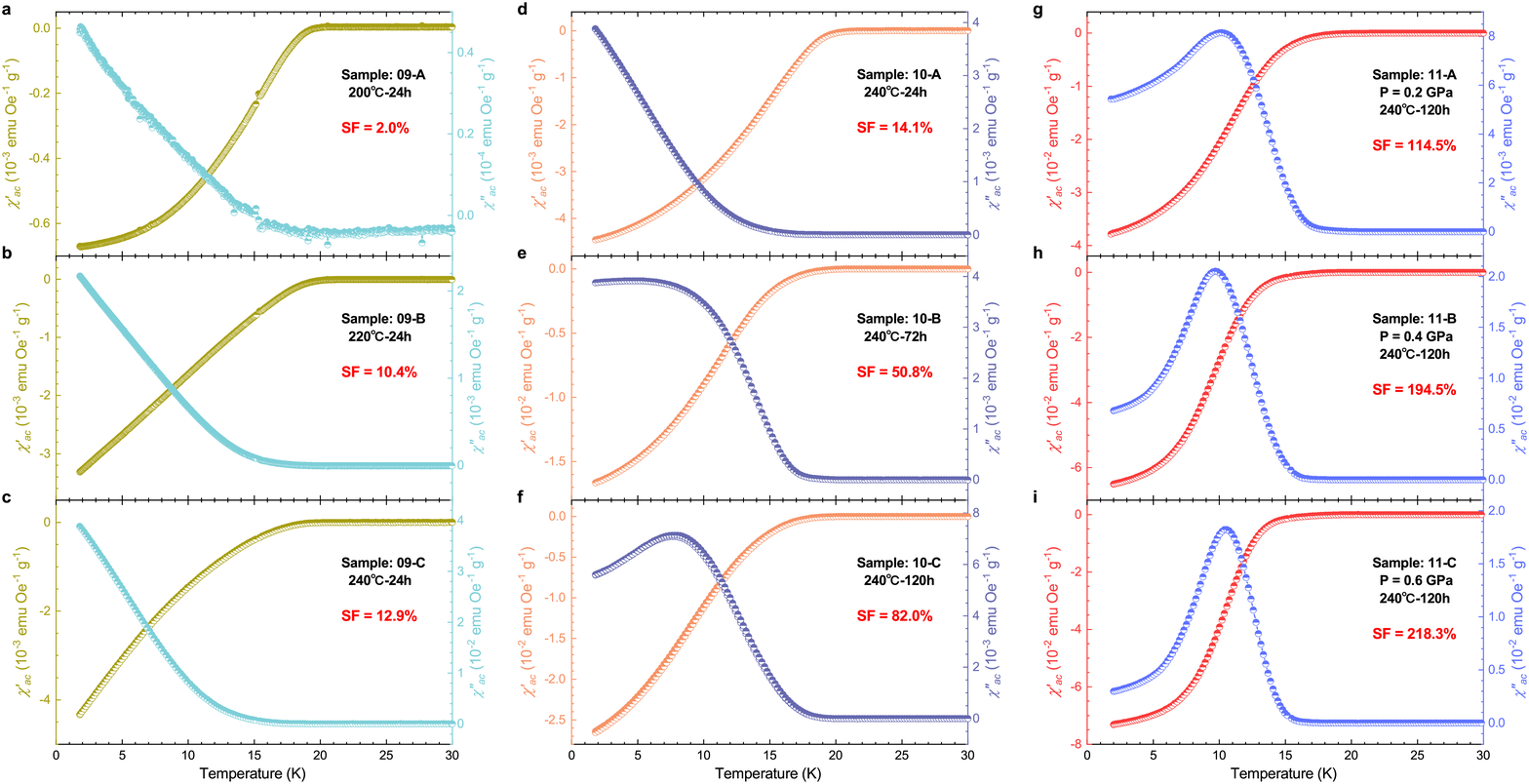}
\end{center}
\vspace{-1.3cm}
\noindent\textbf{Extended Data Figure 3 $\mid$ Temperature dependence of the real part ($\chi^{\prime}_{ac}$) and imaginary component ($\chi^{\prime\prime}_{ac}$) of the \emph{ac} magnetic susceptibility for samples at different synthesis conditions to have the different superconducting shielding fraction (SF).} \textbf{a}, \textbf{b} and \textbf{c}, Temperature dependence of $\chi^{\prime}_{ac}$ and $\chi^{\prime\prime}_{ac}$ of sample No. 9. Parts A, B and C of sample No. 9 were annealed at different temperatures of 200 $^\circ$C, 220 $^\circ$C and 240 $^\circ$C for 24 hours, respectively. $\chi^{\prime}_{ac}$ is essentially close to the \emph{dc} magnetic susceptibility in the ZFC run for the different SFs. The $\chi^{\prime\prime}_{ac}$ value increases with increasing SF. \textbf{d}, \textbf{e} and \textbf{f}, Parts A, B and C of sample No. 10 were annealed at 240 $^\circ$C for the periods ranging from 1 to 5 days, respectively. The imaginary part of the \emph{ac} magnetic susceptibility of the sample with a SF of 50.8 $\%$ appears as a plateau region at low temperatures, and the 82.0 $\%$ of the sample shows a distinct hump. This is characteristic of samples with increasing SF. \textbf{g}, \textbf{h} and \textbf{i}, Parts A, B and C of sample No. 11 were annealed at 240 $^\circ$C for 5 days after the powder samples were pre-pressed into solid cylinders using different pressures. The sharper the peak of $\chi^{\prime\prime}_{ac}$ in higher SF samples. SF is positively correlated with the volume fraction of the superconducting phase in the sample, and the actual measured values are also affected by the non-superconducting fraction of the sample. The frequency and amplitude of the applied field for all \emph{ac} magnetic susceptibility measurements are 233.6 Hz and 4 Oe, respectively.

\newpage
\begin{center}
\includegraphics[width=0.95\columnwidth]{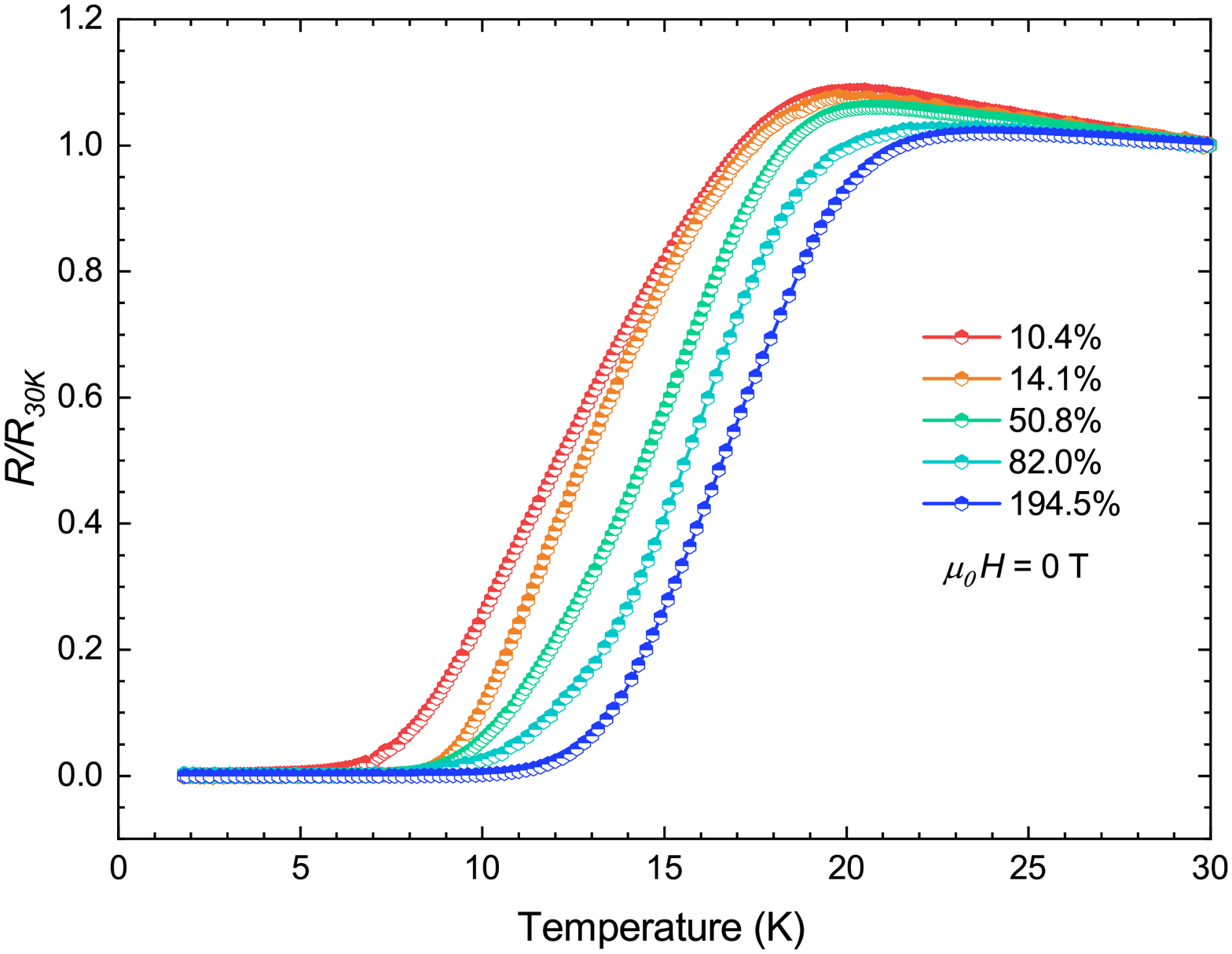}
\end{center}
\vspace{-1.3cm}
\noindent\textbf{Extended Data Figure 4 $\mid$ Temperature dependence of normalized resistance to 30 K (R/R$_{30}$) for samples with different superconducting shielding fractions (SFs) collected in the warming run.} The resistance of the sample increases with decreasing temperature before dropping to the zero value in the superconducting state. Superconducting transition width decreases with increasing SF from 10.4 $\%$ to 194.5 $\%$. Increasing SF leads to the increase of both the on-set temperature for the superconducting transition and the temperature for the sample to enter completely the zero-resistance state. 

\newpage
\begin{center}
\includegraphics[width=0.95\columnwidth]{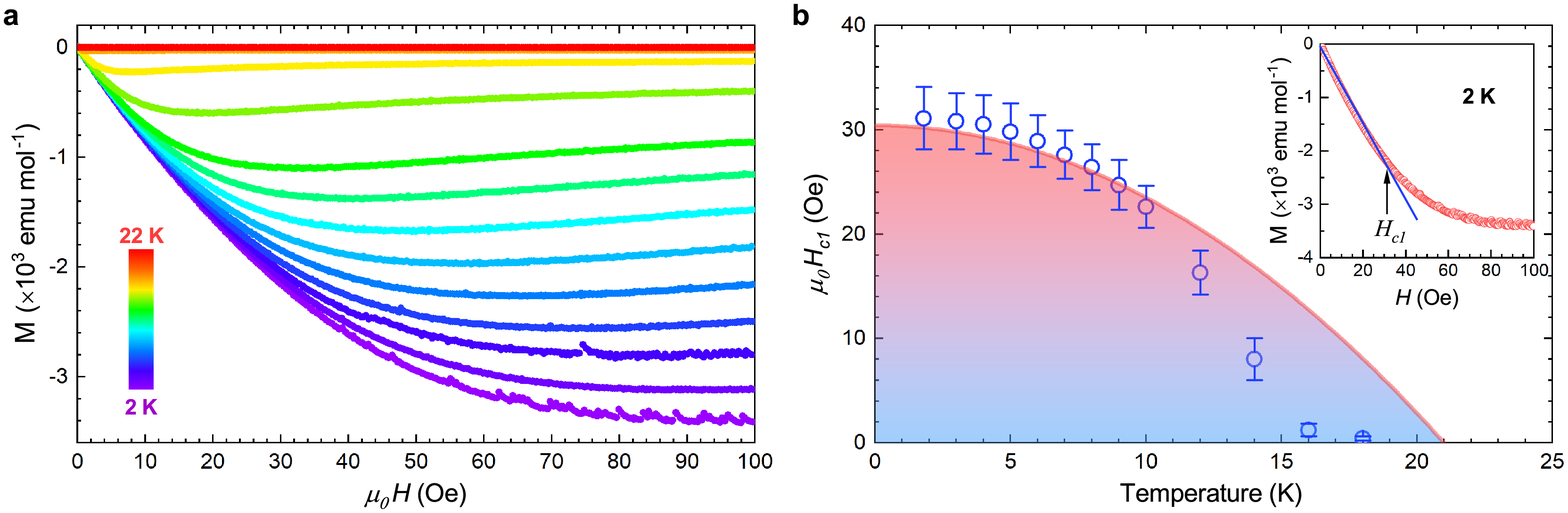}
\end{center}
\vspace{-1.3cm}
\noindent\textbf{Extended Data Figure 5 $\mid$ The lower critical field \emph{H$_{c1}$} of Cs$_3$C$_{60}$.} \textbf{a}, Magnetic field dependence of the magnetization at various temperatures from the superconducting state to the normal state. The temperature step of 1 K below 10 K, and 2 K above 10 K in the measurements. \textbf{b}, Temperature dependence of the lower critical field \emph{H$_{c1}$}. The error bars represent estimated uncertainties in determining \emph{H$_{c1}$}. The inset at the upper right shows the determination for \emph{H$_{c1}$} at the temperature of 2 K. The solid line of shadow border represents the empirical law \emph{H$_{c1}$}(T)/\emph{H$_{c1}$}(0) = 1 - (\emph{T}/\emph{T}$_c$)$^2$. The fitting to the measured values of the lower critical magnetic field near zero temperature yields \emph{H$_{c1}$}(0). The \emph{H$_{c1}$} values at temperatures higher than 10 K do not affect much the fitting accuracy of \emph{H$_{c1}$}(0).

\newpage
\begin{center}
\includegraphics[width=0.95\columnwidth]{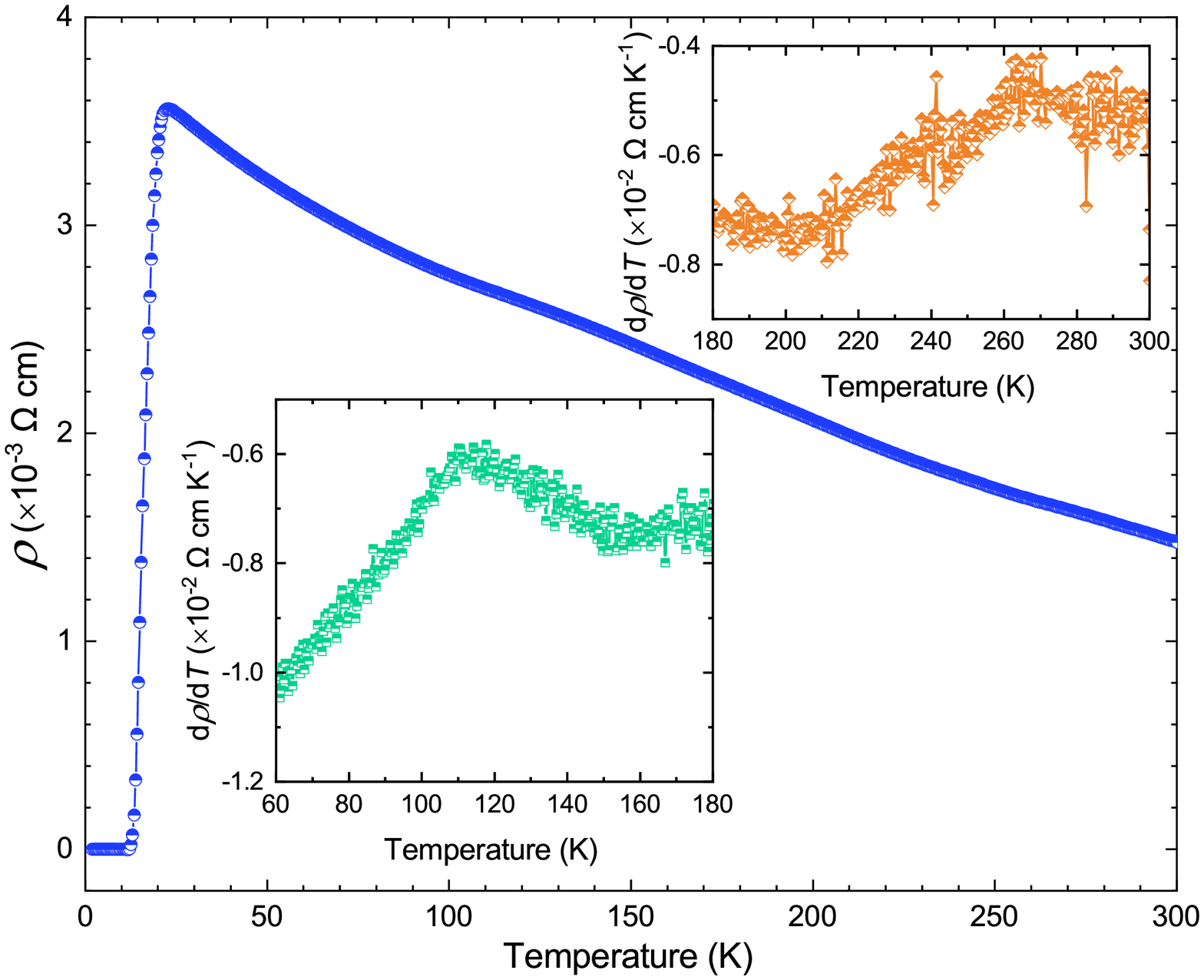}
\end{center}
\vspace{-1.3cm}
\noindent\textbf{Extended Data Figure 6 $\mid$ Temperature dependence of the resistivity of Cs$_3$C$_{60}$ in the warming run.}  The resistivity of the sample increases with decreasing temperature before dropping to the zero value in the superconducting state. Two insets: The temperature dependence of the derivative of the resistivity ($d\rho$/$dT$) in two different temperature regimes. Two anomalies can be observed at around 110 K and 250 K, respectively.

\end{document}